\documentclass[sigconf]{acmart}

\usepackage{booktabs} % For formal tables
\usepackage{color}
\usepackage[english]{babel}
\usepackage{array}
\usepackage{bm}
\usepackage{multirow}
\usepackage{enumitem}
\usepackage{pifont}

\usepackage{flushend}

\setcopyright{rightsretained}
\acmDOI{10.475/123_4}

\acmISBN{123-4567-24-567/08/06}

\acmConference[DAC]{}{2019}{Las Vegas}
\acmYear{2019}
\copyrightyear{2019}

\acmArticle{4}
\acmPrice{15.00}

\editor{Jennifer B. Sartor}
\editor{Theo D'Hondt}
\editor{Wolfgang De Meuter}

\begin{document}
%\title{FNAS: A Generic Framework for Optimizing DNNs on FPGAs via Fine-Grained Scheduling}
%\title{FNAS: A Novel Framework to Accelerate Neural Architecture Search for FPGA-based CNN Inference}
\title{Accuracy vs. Efficiency: Achieving Both through FPGA-Implementation Aware Neural Architecture Search}

\author{{
Weiwen Jiang$^{1,2,3}$ \quad
Xinyi Zhang$^{2}$ \quad
Edwin H.-M. Sha$^{1}$ \quad
Lei Yang$^{3,4}$ \quad
Qingfeng Zhuge$^{1}$ \quad\quad
Yiyu Shi$^{5}$ \quad
Jingtong Hu$^{2}$
}
{
\normalsize
\\
$^{1}$ East China Normal University
$^{2}$ University of Pittsburgh
$^{3}$ Chongqing University\\
$^{4}$ University of California, Irvine
$^{5}$ University of Notre Dame
\\
jiang.wwen@pitt.edu
}
\vspace{10pt}
}

\begin{abstract}
A fundamental question lies in almost every application of deep neural networks: what is the optimal neural architecture given a specific dataset? Recently, several Neural Architecture Search (NAS) frameworks have been developed that use reinforcement learning and evolutionary algorithm to search for the solution. However, most of them take a long time to find the optimal architecture due to the huge search space and the lengthy training process needed to evaluate each candidate. In addition, most of them aim at accuracy only and do not take into consideration the hardware that will be used to implement the architecture. This will potentially lead to excessive latencies beyond specifications, rendering the resulting architectures useless. To address both issues, in this paper we use Field Programmable Gate Arrays (FPGAs) as a vehicle to present a novel hardware-aware NAS framework, namely {\em FNAS}, which will provide an optimal neural architecture with latency guaranteed to meet the specification. In addition, with a performance abstraction model to analyze the latency of neural architectures without training, our framework can quickly prune architectures that do not satisfy the specification, leading to higher efficiency.
Experimental results on common data set such as ImageNet show that in the cases where the state-of-the-art generates architectures with latencies $7.81\times$ longer than the specification, those from FNAS can meet the specs with less than $1\%$ accuracy loss. Moreover, FNAS also achieves up to $11.13\times$ speedup for the search process. To the best of the authors' knowledge, this is the very first hardware aware NAS.
\end{abstract}

\maketitle

\setlength{\textfloatsep}{5pt}
\setlength{\floatsep}{5pt}
\setlength{\dbltextfloatsep}{5pt}

\section{Introduction}

The performance of a Deep Neural Network (DNN) is mostly decided by its architecture. Yet the design of DNN architecture had significantly relied on human expertise and labor until the recent development of Neural Architecture Search (NAS) that can automatically explore the optimal architecture for a particular application.
Existing research efforts \cite{zoph2016neural,real2017large} have demonstrated that NAS can generate DNNs of competitive or even better accuracy against the human-invented ones (e.g., AlexNet, VGGNet, GoogleNet and ResNet). However, the prevalence of NAS is obstructed by its efficiency. As reported in \cite{zoph2016neural}, the search process can take several days even with hundreds of GPUs. The issue mainly comes from the fact that the search space can be huge, and for each candidate architecture, lengthy training process is needed to evaluate it.

In addition, a mainstay for any existing NAS framework is that accuracy is the mono-objective to guide the search \cite{zoph2016neural}. If the resulting architecture is to be deployed in the cloud or latency is not a critical factor, they will still work. However, if the architecture is to be implemented on hardware with latency specification, then there is no guarantee that the specification will be met. In these scenarios, the optimal architecture found by NAS is simply useless.

In this paper, we propose a novel hardware-aware NAS framework to address the above issues. To illustrate our framework, we choose to use Field programmable gate array (FPGA) as a vehicle, as it has gradually become one of the most popular platforms to implement DNNs due to its high performance and energy efficiency, in particular for low-batch real-time applications \cite{chung2018serving}. To introduce hardware awareness, it seems to be straightforward to simply include an additional metric in existing NAS frameworks that describe the latency of a neural architecture on an FPGA. However, the evaluation of the metric can be challenging. First, unlike the regular path-based structures in most human-invented DNNs, the architecture obtained by NAS can be irregular. Many existing design flows dedicated for human-invented DNNs are not suitable for such complicated structures \cite{zhang2015optimizing,shen2017maximizing,zhang2018dnnbuilder,wei2018tgpa,zhang2016energy,jiang2018heterogeneous,chung2018serving,fowers2018configurable}.
Second, the architectures from NAS commonly have large sizes, which may require multi-FPGAs to collaborate for implementation.
Consequently, the scheduling of tasks on multiple FPGAs should be taken into consideration. As such, a more elegant way to evaluate the metric is warranted.

Towards this, we put forward an abstraction model that builds a bridge between the software (neural architecture) and hardware (FPGA designs) for efficient latency estimation.
Specifically, a tile-based graph model is presented to describe a given DNN under an FPGA design.
In the model, we determine the granularity of tasks, the task dependencies, and the data accesses according to the DNN architecture and the tiling parameters in the design.
Then, the complicated dependencies among DNN layers can be captured by adding extra edges among tasks.

As for the scheduling on multiple FPGAs, a limited number of works exist in the literature \cite{zhang2016energy,jiang2018heterogeneous}, all of which still follow the scheduler design for a single FPGA \cite{zhang2015optimizing} (see Figure \ref{Fig:rwFPGAImpl}(a)).
Such schedule paradigm, however, cannot fully exploit the parallelism among FPGAs.
In this work, we propose a more flexible schedule mechanism (see Figure \ref{Fig:rwFPGAImpl}(b)). We first study the design principle for schedulers, based on which we present the
mechanism to schedule tasks in the abstraction model.
Furthermore, we theoretically analyze the latency of executions and pipeline stalls to estimate the overall latency.
Kindly note that the proposed schedule paradigm can also be widely applicable in the design of multi-FPGA systems beyond the scope of this work.

The main contributions of this paper are as follows.
\begin{itemize}
  \item \textbf{Framework.} We build an FPGA-implementation aware neural architecture search framework, namely {\em FNAS}, which can generate optimal DNN architectures with guaranteed latency on target FPGAs.
  \item \textbf{Abstraction Model.} We propose a graph model to describe neural architectures based on FPGA implementations, which provides the fundamental support for latency analysis. In addition, it can model different kinds of architectures.
   \item \textbf{Schedule Paradigm.} We present a novel schedule paradigm, which can fully exploit the parallelism among multiple FPGAs.
\end{itemize}

Experimental results on common data set such as ImageNet show that in the cases where the state-of-the-art \cite{zoph2016neural} generates architectures with latencies $7.81\times$ longer than the specifications, those from FNAS can meet them with less than $1\%$ accuracy loss; meanwhile FNAS also achieves $11.13\times$ speedup for the search process.

The remainder of the paper is organized as follows. Section 2 reviews related background and Section 3 demonstrates our motivation and problem formulation. The detailed FNAS algorithm is presented in Section 4. Experimental results are shown in Section 5 and concluding remarks are given in Section 6.

\section{Background}\label{Sec:mot}

\begin{figure}[t]
  \centering
  \includegraphics[width=3.45in]{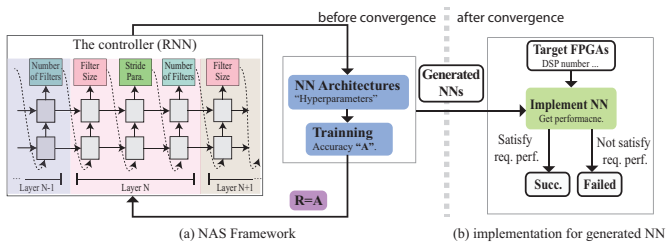}
  \vspace{-10pt}
  \caption{NAS framework \cite{zoph2016neural} with its implementation.}\label{Fig:NASOriFramework}
\end{figure}

In this section, we will present the background on the neural architecture search and the FPGA-based DNN implementations.

\textbf{Searching Neural Network Architecture.} Although the research on automatically predicting neural network architectures can trace back to the 1980s \cite{schaffer1992combinations}, after deep neural networks have achieved great success in AI domains, there has been growing interests in generating good neural architectures recently.
With the fact that the architectures are growing deeper, the search space grows exponentially, which makes the search process difficult.
In existing works, there are two main directions in searching an architecture: (1) employing reinforcement learning \cite{zoph2017learning,zoph2016neural,baker2016designing}, and (2) applying the evolutionary algorithms \cite{real2017large,xie2017genetic}.
Figure~\ref{Fig:NASOriFramework} shows the NAS framework presented in \cite{zoph2016neural}.
In NAS, it iteratively generates a new child network, and obtains its accuracy $A$ by training it on a held-out data set.
Then, accuracy $A$ will be used as the reward signal for the next iteration.
The search process will be stopped if the controller is converged for the maximum accuracy, or the accuracy of child network satisfies the required accuracy $rA$. The generated final design will then be implemented into FPGAs. Existing work has demonstrated that the automatically searched network architectures can achieve close accuracy to the best human-invented architectures \cite{zoph2017learning,zoph2016neural}.
However, there are two important challenges that need to be addressed.
First, the searching process is inefficent. \cite{zoph2016neural} reported that 20,000 networks were trained across 500 P100 GPUs over 4 days to find the descired network.
Second, the generated neural architectures achieve high accuracy with the sacrifice of inference speed.
The resultant network are usually complex and slow, which frequently fails to satisfy the required timing specification with available computing resources for real-time AI applications.

Table \ref{Tab:MotComp} reports our results of NAS \cite{zoph2016neural} for image classification using MNIST data set targeting on PYNQ board \cite{pynq2018}.
We can observe that it takes 190\emph{m} 33\emph{s} to complete the search by NAS, with 19.70 \emph{ms} of the latency and 99.42\% of the accuracy for the generated network.

\begin{table}
  \centering
  \tabcolsep 5.7pt
  \renewcommand\arraystretch{1.3}
  \footnotesize
  \caption{FNAS uses less time to generate architectures having lower latency on PYNQ with small accuracy degradations.}
    \vspace{-10pt}
        % Table generated by Excel2LaTeX from sheet 'Sheet2'
        \begin{tabular}{|c|c|cc|cc|cc|}
        \hline
        \multirow{2}{*}{Methods} &  TC  & \multicolumn{ 2}{c|}{Elasp.} & \multicolumn{ 2}{c|}{Lat.} & \multicolumn{ 2}{c|}{Acc.} \\
        \cline{2-8}
         &            \emph{ms} &   (m,s) &       Imp. &       (ms) &       Imp. & (\%) &       Deg. \\
        \hline
               NAS \cite{zoph2017learning} & -  &        190m33s &          - &       19.70 &          - & 99.42\% &          - \\
        \hline

              \multirow{3}{*}{\textbf{FNAS}}  &  10  &  \textbf{74m29s} &   \textbf{2.55$\bf{\times}$}         &      \textbf{8.67} &  \textbf{2.27$\bf{\times}$} &     \textbf{99.34\%} &    \textbf{-0.08\%}      \\
                                              &  5  &  \textbf{59m19s} &    \textbf{3.21$\bf{\times}$}        &      \textbf{4.77} &
              \textbf{4.13$\bf{\times}$} &     \textbf{99.18\%} &    \textbf{-0.24\%}   \\
                                              &  2  &  \textbf{17m07s} &    \textbf{11.13$\bf{\times}$}        &      \textbf{1.80} &
              \textbf{10.94$\bf{\times}$} &     \textbf{98.61\%} &    \textbf{-0.81\%}  \\
        \hline
        \end{tabular}
\label{Tab:MotComp}
\end{table}

\textbf{FPGA Implementation.} FPGA has demonstrated its excellent abilities to achieve high performance and energy efficiency for low-batch real-time inferences. With such vision, a series of works for implementing neural networks on FPGAs have been carried out: (1) a single accelerator is implemented on a single FPGA \cite{zhang2015optimizing,xu2018resource}; (2) multiple accelerators are integrated into a single FPGA \cite{shen2017maximizing,zhang2018dnnbuilder,wei2018tgpa}; (3) multiple accelerators are deployed on multiple FPGAs \cite{zhang2016energy,jiang2018heterogeneous,chung2018serving,fowers2018configurable}. Existing NAS did not take implementation into consideration at all. In order to make NAS process implementation-aware, the inference latency in FPGA has to be obtained for each child network. Existing research efforts commonly analyze the latency of DNNs on FPGAs by generating the HLS- or RTL-level code \cite{zhang2015optimizing,shen2017maximizing,zhang2018dnnbuilder}, which may involve human intervention and vast amount of time. Therefore, if we simply obtain inference latency using existing techniques and naively integrating it into the reward for architecture selection, the NAS search process will take even longer time.

In this work, one of the main contributions is to accurately and quickly estimate FPGA inference latency so that both search process and network generating are efficient. Table \ref{Tab:MotComp} also reports results of the proposed FNAS. From this table, we can see that for the same dataset, by setting the inference timing specifications (TS) to 2\emph{ms}, FNAS can reduce the search time from 190 minutes to 17 minutes, achieving 11.13$\times$ reduction. What is more, the latency of the resultant architecture on PYNQ is $10.94\times$ shorter against that explored by NAS. Meanwhile, the accuracy degradation is within 1\%.
For other two cases, FNAS significantly achieves more than 2$\times$ speedup with only 0.08\% and 0.24\% penalty on accuracy.
These verify that the implementation of aware-FNAS can significantly reduce the search time and guarantee the generated architecture on target FPGAs to satisfy the  timing specification in the inference phase while maintaining the accuracy.

\section{FNAS Framework}\label{Sec:Framework}

\vspace{2pt}
\noindent\textbf{\Large 3.1 Problem Definition and FNAS Overview}

In this paper, we aim to develop an FPGA-implementation aware Neural Architecture Search.
The problem is formally defined as follows: Given a specific data set, a target FPGA platform and a required inference latency $rL$, our objective is to automatically generate a neural network, such that its inference latency on the given FPGA platform is less than $rL$, while achieving the maximum accuracy for the machine learning task on the given data set.

\begin{figure}[t]
  \centering
  \includegraphics[width=3.24in]{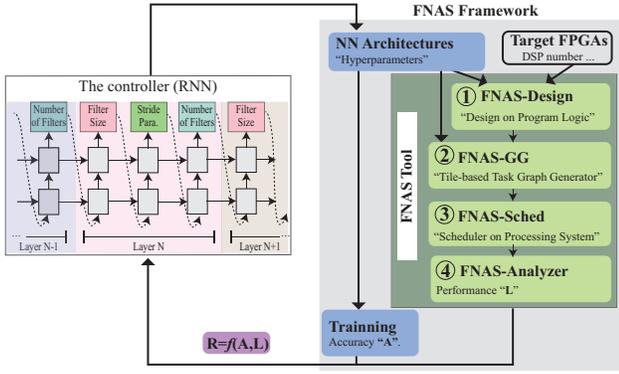}
  \vspace{-10pt}
  \caption{Overview of the proposed FNAS framework.}\label{Fig:Framework}
\end{figure}

Figure \ref{Fig:Framework} shows the overview of FPGA-implementation aware Neural Architecture Search (FNAS) framwork.
In FNAS, it takes the FPGA-based inference performance into consideration during child network searching.
Specifically, instead of directly applying accuracy $A$ as reward, FNAS employs a reward function $f$ to calculate the reward in terms of accuracy $A$ and performance/latency $L$.
In order to efficiently and accurately estimate the inference latency $L$ for a given NN architecture on target FPGAs, we develop the ``FNAS tool''. There are 4 components in FNAS tool, including {\large\ding{192}} \textbf{FNAS-Design}, {\large\ding{193}} \textbf{FNAS-GG}, {\large\ding{194}} \textbf{FNAS-Sched}, and {\large\ding{195}} \textbf{FNAS-Analyzer}.
In the following texts, we will first present the reward function, then introduce these components one-by-one.

\vspace{2pt}
\noindent\textbf{\Large 3.2 Reward function}

Reward function takes the accuracy $A$, latency $L$, and the required latency $rL$ to calculate the reward signal.
The function to calculate the reward $R$ is defined as follows.
\begin{equation}\label{equ:reward}
R = \left\{ {\begin{array}{*{20}{c}}
{\frac{{rL - L}}{{rL}} - 1}&{L > rL}\\
{\left( {A - b} \right) + \frac{{L}}{{rL}}}&{L \le rL}
\end{array}} \right.
\end{equation}
In the above function, there are two cases.
First, if $L>rL$, it indicates that the performance of the resultant system cannot satisfy the timing specification.
In this case, we do not train the child network and directly return a negative reward to the controller.
In the second case, we sum up the reward of performance and accuracy, where the performance reward is set as $\frac{{L}}{{rL}}$, which indicates a solution has higher performance reward if its latency approaches the required level.
Here, $b$ is a baseline function, which is an exponential moving average of the previous architecture accuracies \cite{zoph2016neural}.

\begin{figure}[t]
  \centering
  \includegraphics[width=3.3397 in]{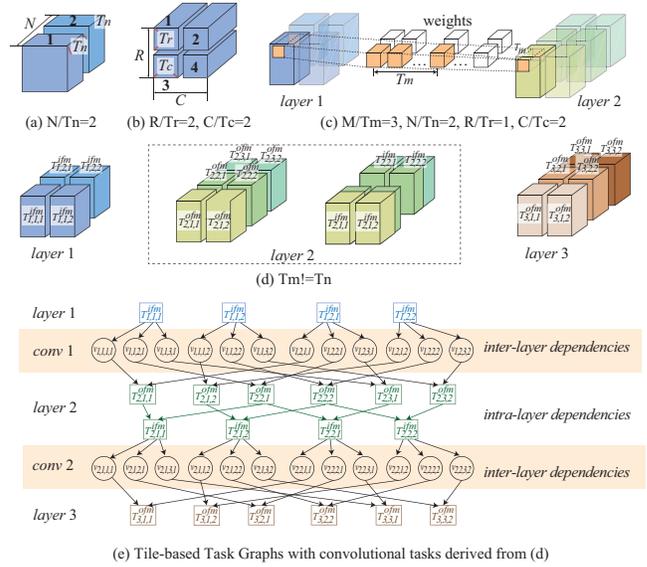}
  \vspace{-10pt}
  \caption{FNAS-Design and FNAS-GG: (a) channel tiles; (b) row/col tiles; (c) tile-based convolution; (d) encoding of tiles; (e) tile-based task graph.}\label{Fig:GGTile}
\end{figure}

\vspace{2pt}
\noindent\textbf{\Large 3.3 Tiling Parameters: {\large\ding{192}} FNAS-Design}

Due to the limited resource on FPGA, it may be difficult to place a whole convolutional layer on FPGA.
In consequence, it is common to apply tiling technique to split convolutional operations into multiple small tasks \cite{zhang2015optimizing,shen2017maximizing,zhang2016energy,zhang2018dnnbuilder,yang2018optimal}.
FNAS-Design is to determine the tiling parameters for a given NN architecture on target FPGAs.

Take one convolutional operation as an example, it involves four parameters $\langle T_m,T_n,T_r,T_c\rangle$, related to the input/output feature maps (IFM/OFM).
Here, the number of IFM channel is $N$. The size of corresponding tiles is $T_n$ (channels). IFM is then partitioned into $\lceil \frac{N}{T_n}\rceil$ tiles, as shown in Figure \ref{Fig:GGTile}(a).
Similarly, OFM with $M$ channels is partitioned to $\lceil \frac{M}{T_m}\rceil$ tiles.
In addition, the numbers of row/column of OFM are $R$ and $C$, respectively. They are tiled according to $T_r$ and $T_c$ as shown in Figure \ref{Fig:GGTile}(b).

After tiling the IFM/OFM/row/col, one convolutional operation is divided to smaller tasks, as shown in Figure \ref{Fig:GGTile}(c).
Each task corresponds to a pair of IFM/OFM tiles.
Tasks in one layer will be continuously loaded to a Processing Element (PE) on FPGA for execution (the load sequence is determined by {\large\ding{194}} FNAS-Sched).
For a task, it involves $K_h\times K_w\times T_r\times T_c\times T_m\times T_n$ Multiply-Accumulate (MAC) operations, where $K_h$ and $K_w$ are the height and weight of filter determined by the controller.
A PE composed of $T_m\times T_n$ DSPs can execute $T_m\times T_n$ MAC operations (16bit fix-point) in parallel \cite{zhang2015optimizing}.
Then the latency of a task is $K_h\times K_w\times T_r\times T_c$.

In FNAS, each layer is allocated to a dedicated PE, and PEs are performed in the pipeline fashion.
Such architecture can be implemented on one FPGA as in \cite{shen2017maximizing,zhang2018dnnbuilder} or multiple FPGAs as in \cite{zhang2016energy,jiang2018heterogeneous}.
The resource (e.g., DSP and memory bandwidth) for each layer can be obtained by considering the load balance.
And then the best parameters $\langle T_m,T_n,T_r,T_c\rangle$ can be obtained according to \cite{zhang2015optimizing,shen2017maximizing}.

\vspace{2pt}
\noindent\textbf{\Large 3.4 Tile-based task graph generator:  {\large\ding{193}} FNAS-GG}

FNAS-GG is a graph generator that takes the design parameters and NN architecture to generate the dependency graph between data tiles and tasks, called tile-based task graph.
To generate the graph, the generator first needs to define each tile for a given design. Then, it can generate the tile-based task graph.

According to FNAS-Design, there are two kinds of tiles: channel tile and row/col tile.
For channel tile, let $CH^{ifm}_i=\{1,2,\cdots,\lceil \frac{CH_i}{T_n}\rceil\}$ be a set of indices in the $i^{th}$ layer under tiling parameter $T_n$ (considering $i^{th}$ layer's IFM); similarly $CH^{ofm}_i=\{1,2,\cdots,\lceil \frac{CH_i}{T_m}\rceil\}$ is under tiling parameter $T_m$ (considering $i^{th}$ layer's OFM).
For row/col tile, let $RC_i=\{1,2,\cdots,\lceil\frac{R_i}{T_r}\rceil \cdot \lceil\frac{C_i}{T_c}\rceil\}$ be a set of indices in layer $i$ under tiling parameter $T_r$ and $T_c$.

Based on these parameters, we can define the tiles as follows.
We define a tile in IFM as $T^{ifm}_{i,j,m}$, indicating the tile is in the $i^{th}$ layer, the $j^{th}$ channel tile in $CH^{ifm}_i$ and $m^{th}$ row/col tile in $RC_i$.
Similarly, the tile in OFM is defined as $T^{ofm}_{i,k,m}$, where $k$ is the index of channel tile in $CH^{ofm}_i$.

Then, we can build the dependencies among data tiles and tasks.
There are two kinds of dependencies.
First, \emph{inter-layer dependencies} describe the dependency between tasks and data tiles in two consecutive layers.
We define a task node to be $v_{i,j,k,m}$, which process the tile $T^{ifm}_{i,j,m}$ and generate the tile $T^{ofm}_{i+1,k,m}$.

Next, the \emph{intra-layer dependencies} describe dependency between two data tiles in one layer.
If the tiling parameters $T_m$ and $T_n$ for a layer are different, as shown in Layer 2 in Figure \ref{Fig:GGTile}(d), the dependency would not be the simple one-to-one mapping
Instead, it can be represented as follows.
For the tile $T^{ifm}_{i,j,m}$, its data is depending on tile $T^{ofm}_{i,k,m}$ if $(j-1)\cdot\frac{Tn}{Tm}+1 \le k\le  j\cdot\frac{Tn}{Tm}$, where $k\in\mathbb{N}$.

By following the above rules, the graph can be generated.
For the tiles of three layers in Figure \ref{Fig:GGTile}(d), its corresponding tile-based task graph is shown in Figure \ref{Fig:GGTile}(e).

\begin{figure}[t]
  \centering
  \includegraphics[width=3.3 in]{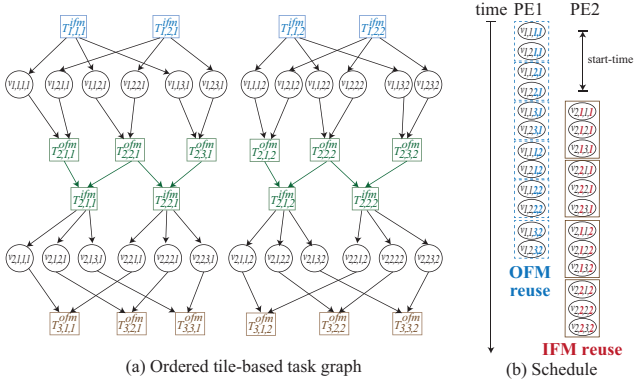}
  \vspace{-10pt}
  \caption{FNAS-Sched: (a) re-ordered graph from \ref{Fig:GGTile}(e); (b) the generated schedule graph.}\label{Fig:GGSched}
\end{figure}

\vspace{2pt}
\noindent\textbf{\Large 3.5 Scheduler design: {\large\ding{194}} \textbf{FNAS-Sched}}

FNAS-Sched is a scheduler to determine the sequence of tasks to be executed on multiple PEs, such that the schedule length (latency) can be minimized.
FNAS-Sched tries to maximally exploit the parallelism among different convolutional operations based on the tile-based task graph.
The design follows three principles:
\begin{itemize}[noitemsep,topsep=0pt,parsep=0pt,partopsep=0pt]
  \item $\mathcal{P}1$. Minimizing the start time of each PE to execute tasks as early as possible.
  \item $\mathcal{P}2$. Maximizing the reuse of data on FPGA to reduce the communication bandwidth requirement.
  \item $\mathcal{P}3$. Minimizing the pipeline stall caused by the lack of input data in the execution on one PE.
\end{itemize}

FNAS-Sched is carried out in three steps.

\emph{Step 1: determine the sequence of IFM tiles in each layer - $\mathcal{P}1$.}
The order of executing IFM will affect the start time of the next layer. There are two possible strategies: \emph{i)} increase the indices of channel tiles first; or \emph{ii)} increase the indices of row/col tiles first.
Because an OFM tile is related to all IFM channels, strategy \emph{i)} is more favored than \emph{ii)} to make the next layer start earlier.

\emph{Step 2: determine the sequence of OFM tiles in each layer - $\mathcal{P}1$.}
The sequence of OFM tiles will determine the ready time of IFM tiles in the same layer.
After Step 1, the sequence of IFM tiles has already been determined.
Hence, we sequentially visit IFM tile in a layer and arrange the OFM tiles that are dependent on it.

\emph{Step 3: determine the task sequence - $\mathcal{P}2$,$\mathcal{P}3$.}
Task sequence will determine the data reuse rate.
There are two data reuse strategies can be exploited: \emph{i)} OFM reuse, and \emph{ii)} IFM reuse.
The OFM (or IFM) reuse indicates that the consecutively executed tasks have the same OFM (or IFM) tile and the same row/col tile.
We observe the uniform reuse strategy for all layers will lead to the pipeline stall due to the lack of input data.
In FNAS, we will alternatively apply the above two strategies for consecutive layers.

Followed by the above three steps, we can obtain a schedule of tasks.
For the tile-based task graph in Figure \ref{Fig:GGTile}(e), Figure \ref{Fig:GGSched}(a) gives the graph with the reordered IFM/OFM tiles.
We also give the schedule for this graph in Figure \ref{Fig:GGSched}(b).
As shown in Figure \ref{Fig:GGSched}(b), tasks in layer1 (PE1) can achieve OFM reuse, while IFM reuse can be achieved in layer2 (PE2).
In addition, the start-time is only 4 time units, and there is no stall in the executions for both layers.

\begin{figure}[t]
  \centering
  \includegraphics[width=3.2732 in]{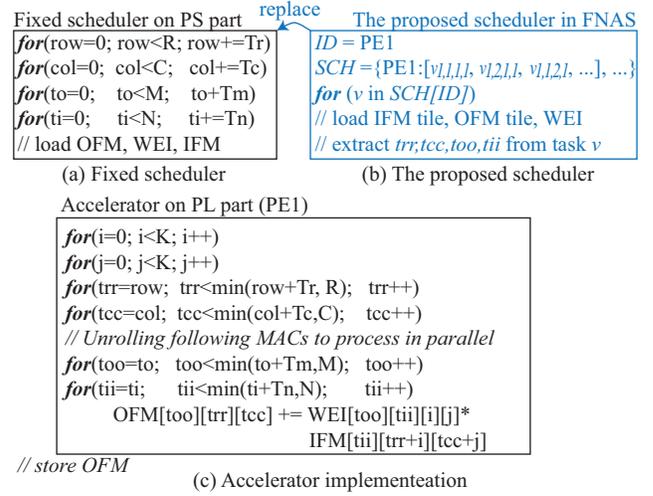}
  \vspace{-10pt}
  \caption{FPGA-based CNN implementation.}\label{Fig:rwFPGAImpl}
\end{figure}

In a FPGA implementation, the scheduler is usually implemented in the processing system (PS) end and the accelerator is implemented on the programming logics (PL) end. Figures \ref{Fig:rwFPGAImpl}(a),(c) illustrate the commonly used PS/PL designs proposed by \cite{zhang2015optimizing}.
In their design, different data tiles are sent to PL part with a fixed order, called ``fixed scheduling''.
In order to implement the proposed scheduler, we modify the implementation of PS part, as shown in Figure \ref{Fig:rwFPGAImpl}(b), where we can specify the tasks to PEs and launch tasks in an order determined by our scheduler.

\vspace{2pt}
\noindent\textbf{\Large 3.6 Latency analysis: {\large\ding{195}} \textbf{FNAS-Analyzer}}

FNAS-Analyzer aims to efficiently and accurately compute the latency $L$ of a neural architecture on target FPGAs with determined schedule.
In the schedule, the latency of PE is the sum of three parts: (1) the processing time, (2) the start time, and (3) the stall time.
We are going to analyze each of them in following sections.

\emph{Processing Time.} We first determine the execution time of tasks in the tile-based graph.
Since all tasks in layer $i$ utilize the same accelerator for execution, they have the same execution time, denoted as $ET_i$ which equals $K_{h,i}\times K_{w,i}\times T_{r,i}\times T_{c,i}$ (see {\large\ding{192}} FNAS-Design).
The processing time $PT_i$ of a PE is the summation of execution time of all task nodes in the corresponding layer $i$, which can be directly calculated as follows.
\begin{equation}
PT_i = ET_i\times |CH_i^{ifm}| \times |CH_{i+1}^{ofm}|
\end{equation}
where $|CH_i^{ifm}|\times |CH_{i+1}^{ofm}|$ is the task number in layer $i$ (see {\large\ding{193}}).

\emph{Start Time.} The start time of a layer depends on the start time and the data reuse strategy of its previous layer.
Let $\Delta t_{i,ofm}$ be the difference between start time of layers $i-1$ and $i$.
We define $\Delta t_{j,ifm}$ similarly.

First, let us consider that layer $i-1$ applies OFM reuse, indicating one tile in OFM is ready after computing $\lceil \frac{CH_{i-1}}{T_{n,{i-1}}}\rceil$ tasks.
For one $IFM$ in layer $i$, it requires $\lceil \frac{T_{n,i}}{T_{m,i}}\rceil$ OFM tiles.
In consequence, $\Delta t_{i,ofm}$ can be calculated as follows.
\begin{equation}
\Delta t_{i,ofm} = \lceil \frac{CH_{i-1}}{T_{n,{i-1}}}\rceil \times \lceil \frac{T_{n,i}}{T_{m,i}}\rceil\times ET_{i-1}
\end{equation}

Next, consider layer $i-1$ applies IFM reuse to compute $\Delta t_{j,ifm}$.
In this case, each IFM in layer $i-1$ will be reused to compute the partial sum of all related OFMs.
After all IFM except the last one in the same row/col tile have completed, it will continuously generate OFMs in layer $i$.
Thus, $\Delta t_{j,ifm}$ can be calculated as follows.
\begin{equation}
\Delta t_{j,ifm} = \left[\left(\lceil \frac{CH_{i-1}}{T_{n,{i-1}}}\rceil-1\right) \times \lceil \frac{CH_{j}}{T_{m,{j}}}\rceil + \lceil \frac{T_{n,i}}{T_{m,i}}\rceil\right]\times ET_{j-1}
\end{equation}
where the first term of multiplication indicates the number of tasks completed before the first OFM in layer $i$ is produced.
And $\lceil \frac{T_{n,i}}{T_{m,i}}\rceil$ is the number of OFM required by one IFM in layer $i$.

\emph{Stall Time.} After a PE has been launched, it may be stalled because the data for the next task is not ready.
However, there may exist another task that has already been ready to run.
In our scheduler, we can maintain a ready-to-run queue.
If a stall occurs, we will pick one task in the ready-to-run queue to avoid pipeline stall.

\emph{Latency.} We can then derive a tight lower bound on latency $Lat_{sys}$ by summing up processing time and starting time.
For a total of $N$ processing elements (PE), assume the first and the last PEs apply OFM reuse, we can calculate $Lat_{sys}$ as follows.
\begin{equation}
Lat_{sys} = \sum\nolimits_{i=2,4,\cdots,N-1}\Delta t_{i,ofm} + \sum\nolimits_{j=3,5,\cdots,N}\Delta t_{j,ifm} + PT_N
\end{equation}
After obtaining $Lat_{sys}$, we have the latency $L=Lat_{sys}$.

\textbf{Summary.} FNAS framework considers the performance of child networks on target FPGAs in the neural architecture search process.
As shown in Formula \ref{equ:reward}, if the latency cannot satisfy the timing specification, there is no need to train the generated child network.
In addition, the controller will be guided to avoid searching architectures that have insufficient performance.
Consequently, the search process can be dramatically accelerated, and the performance of the resultant child network on target FPGAs can be guaranteed.
The evaluations on the efficiency of FNAS will be presented in the next section.

\section{Experimental Results}\label{Sec:exp}
This section will report the evaluation results of the proposed FNAS framework.
Results on MNIST, CIFAR-10, and ImageNet data sets show that FNAS can achieve up to $11.13\times$, $10.89\times$ and $10.38\times$ speedup in the search process respectively.
Moreover, for the case where NAS \cite{zoph2016neural} generates architectures with latencies $9.85\times$ longer than the specification, FNAS can meet the specs with less than 1\% accuracy loss.

\begin{table}
  \centering
  \tabcolsep 1.2pt
  \renewcommand\arraystretch{1.3}
  \footnotesize
  \caption{Data sets and parameter settings in FNAS.}
    \vspace{-10pt}
% Table generated by Excel2LaTeX from sheet 'Sheet2'
    \begin{tabular}{|c|c|c|c|c|c|c|c|c|c|}
    \hline
    \multirow{2}{*}{Data sets} & \multicolumn{ 3}{c|}{Training Para.} &                                                        \multicolumn{ 4}{c|}{Controller Parameters} & \multicolumn{2}{c|}{Para.}\\
    \cline{2-10}
     &      Train &   Val. & E & L & FS & FN & T &    \multicolumn{ 2}{c|}{[TS4,TS3,TS2,TS1]}  \\
     \hline
         \multirow{2}{*}{MNIST} &     \multirow{2}{*}{60,000} &     \multirow{2}{*}{10,000} & \multirow{2}{*}{25} &           \multirow{2}{*}{4} &   \multirow{2}{*}{[5,7,14]} &  \multirow{2}{*}{[9,18,36]} & \multirow{2}{*}{60} &
         TS-High  &  TS-Low \\
          &      &      &  &            &    &   & &
         [2,5,10,20] &  [1,4,10,20] \\

    \hline
       CIFAR-10 &     45,000 &      5,000 &  25 &      10 &  [1,3,5,7] & [24,36,48,64] & 60  &   \multicolumn{2}{c|}{[1.5,2,2.5,10]}  \\
    \hline
      ImageNet &    4,500 & 500 &      25    &         15 &  [1,3,5,7] & [16,32,64,128] & 60  & \multicolumn{2}{c|}{[2.5,5,7.5,10]}         \\
    \hline
    \multicolumn{10}{l}{$\bullet$ L: number of layers; FS: filter size; FN: number of filters; T: trails; E: epoch}

    \end{tabular}
\label{Tab:ExpSet}
\end{table}

\vspace{2pt}
\noindent\textbf{\Large 4.1 Experimental setup}

\textbf{Data sets.} FNAS will search convolutional neural network structures for three kinds of data sets, including MNIST, CIFAR-10, and ImageNet.
Each data set is composed of training set and validation set, as shown in Table \ref{Tab:ExpSet}.
For instance, there are 60,000 and 10,000 examples in training and validation sets, respectively.
Kindly note that for ImageNet, we use a smaller data set to reduce the computation time.
In training of the child networks, the number of epochs is set as 25, and the maximum validation accuracy in the last 5 epochs will be utilized to compute the reward for updating the controller.

\textbf{Controller.} We implement the reinforcement learning based RNN controller based on \cite{zoph2016neural} to generate child networks.
For different data sets, the controller has different configurations as shown in Table \ref{Tab:ExpSet}.
For instance, we explain the configuration of MNIST: (1) its child network has 4 layers, (2) the possible filter size (height and width) is 5, 7 or 14, (3) the possible channel number is 9, 18, or 36, and (4) it will find 60 child networks.

\textbf{FNAS Tool.} We implement all the components described in Section 3 and integrate them into the controller to realize the FNAS framework.
To compare FNAS with NAS, we employ both low-end and high-end FPGAs to implement the resultant architectures using MNIST data set. The low-end and high-end FPGAs selected are Xilinx 7A50T and 7Z020, respectively. To see how general our conclusions are, we have explored CIFAR-10 and ImageNet as additional data sets, where Xilinx ZU9ED is used.

\begin{figure}[t]
  \centering
  \includegraphics[width=3.101 in]{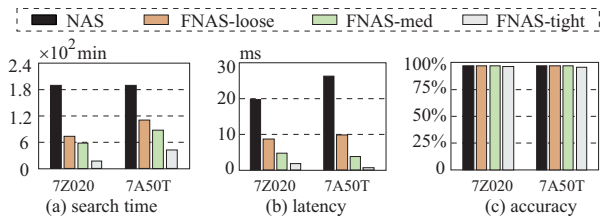}
  \vspace{-10pt}
  \caption{Comparison of search time, latency and accuracy between NAS and FNAS.}\label{Fig:EXPRes}
\end{figure}

\vspace{2pt}
\noindent\textbf{\Large 4.2 Efficiency and accuracy evaluations of FNAS}

Figures \ref{Fig:EXPRes}(a),(b),(c) reports comparison results on search time, latencies of the resultant DNNs, and accuracies of resultant DNNs respectively on the MINST data set.
In these figures, the x-axis represents different FPGAs.
The FNAS-loose (TS2), FNAS-med (TS3), and FNAS-tight (TS4) correspond to three timing specifications (see Table \ref{Tab:ExpSet}).

From Figure \ref{Fig:EXPRes}(a), we can see that FNAS can dramatically reduce the search time.
For the target FPGA of 7Z020, for loose, medium and tight timing specification (actual values in Table \ref{Tab:ExpSet}), the search times are reduced from 190 minutes to 74, 59, 17 minutes,  achieving 2.56$\times$, 3.22$\times$, 11.13$\times$ reductions respectively compared with NAS.
There are two reasons for the improvement: (1) with early-stage pruning, we will not train the architectures whose inference latencies violate the specification; (2) the structure of the valid DNNs are usually simpler than the ones with violations, and accordingly most complex architectures are naturally pruned without the need of training. From the figure we can also see that for FNAS the search time decreases as the timing specification gets tighter, which is also expected as tighter specifications prunes more potential architectures.

Next, as shown in Figure \ref{Fig:EXPRes}(b), for different timing specifications, FNAS can generate a specific architecture to satisfy the specification, i.e., the latency of the architecture decreases as the specification gets tighter.
On the contrary, NAS can only generate a single architecture, which has a latency that is $2.54\times$, $4.19\times$, and $7.81\times$ longer than the specification.
The flexibility of {FNAS} provides more choices for designers and also helps to prune useless designs at the very early design stage.

Figure \ref{Fig:EXPRes}(c) reports the accuracy of the generated architectures.
We can observe that, even compared with the architectures generated by NAS that have timing violations, those generated by FNAS only have accuracy degradation of 0.08\%, 0.24\%, and 0.81\% for loose, medium and tight timing timing specification respectively.
The above results clearly show that FNAS can achieve both high efficiency and high accuracy in exploring neural architectures.

\begin{figure}[t]
  \centering
  \includegraphics[width=3.062 in]{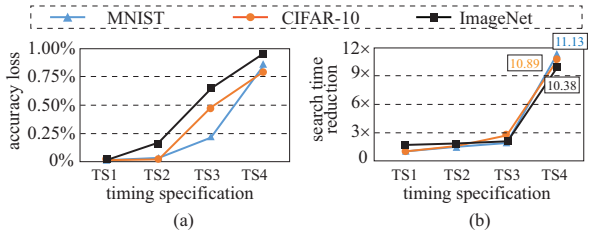}
  \vspace{-10pt}
  \caption{(a) Accuracy loss and (b) search time reduction vs. timing specifications on three data sets. The architectures from NAS are used as the baseline cases.}\label{Fig:EXPTrad}
\end{figure}

Finally, we explore how the accuracy of the architectures generated by FNAS as well as the corresponding search time scales with the timing specifications. The results are depicted in Figure \ref{Fig:EXPTrad} where three data sets MNIST, CIFAR-10, and ImageNet are used. For MNIST, the high-end FPGA is used.
The x-axis represents different timing specifications, where $TC1$ is the loosest one while $TC4$ is the tightest one and the corresponding values are summarized in Table \ref{Tab:ExpSet}. When reporting both accuracy and search time, we use the architecture from NAS as reference and show the accuracy loss as well as the search time reduction.
From Figure \ref{Fig:EXPTrad}(a) we can observe that in general the architectures generated by FNAS without timing violations only have less than 1\% accuracy loss compared with those generated by NAS that have violations. The accuracy loss gets higher as the timing constraint becomes tighter.
Meanwhile, from Figure \ref{Fig:EXPTrad}(b), we can see that the search time of FNAS can be significantly reduced, achieving 11.18$\times$ reduction for CIFAR-10 compared with NAS.

\begin{figure}[t]
  \centering
  \vspace{-5pt}
  \includegraphics[width=3.2732 in]{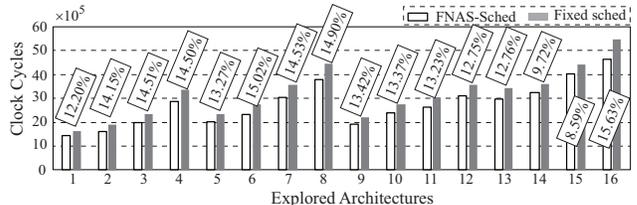}
  \vspace{-13pt}
  \caption{Number of clock cycles on a set of architectures using FNAS-Sched and fixed scheduling \cite{zhang2015optimizing} using four accelerators.}\label{Fig:MotSche}
\end{figure}

\vspace{2pt}
\noindent\textbf{\Large 4.3 Improvements from the scheduler}

We also evaluate the efficiency of the proposed scheduler against that uses the fixed scheduling technique \cite{zhang2015optimizing} on four accelerators.
We test a set of architectures with 4 convolution layers on PYNQ board \cite{pynq2018}.
For each convolution layer, the filter size is $3\times 3$ and the number of filters is 64 or 128.
Figure \ref{Fig:MotSche} reports the number of clock cycles of each possible architecture.
As shown in this figure, our proposed scheduler can consistently achieve better performance.
This convincingly demonstrates the effectiveness of the proposed scheduler for FPGA designs.

\section{Conclusion}\label{Sec:con}

In this work, we use FPGA as a vehicle to explore hardware-aware neural architecture search (NAS).
Our objective is to automatically search the neural architecture with the optimal accuracy while satisfying the timing specification on target FPGAs.
To achieve this goal, a novel FPGA-implementation aware NAS framework is proposed.
In the framework, we build a performance abstraction model and a new schedule paradigm to fully exploit the parallelism among multiple FPGAs.
Evaluation results show that for the cases where the state-of-the-art NAS generates architectures with inference latency $7.81\times$ longer than the specification, the proposed framework can meet the specification with less than 1\% accuracy loss, as well as up to 11.13$\times$ speedup in the search process.

\section*{Acknowledgements}

This work was supported in part by the National Natural Science
Foundation of China under Grant 61472052, in part by the National Science Foundation under Grant CCF-1820537, and in part by the China Scholarship Council under Grant 201706050116 and Grant 201706050117.

\bibliographystyle{ACM-Reference-Format}

\end{document}